\documentclass[aps,prl,twocolumn,email,reprint]{revtex4}

\usepackage{amssymb}
\usepackage{amsmath}
\usepackage{natbib}
\usepackage{graphicx}
\usepackage{bbold}
\usepackage{color}
\usepackage{soul}
\usepackage{pdfpages}

\soulregister\ref{7}
\soulregister\eqref{7}
\soulregister\cite{7}
\soulregister\onlinecite{7}

\begin{document}
	
	\title{Spatio-temporal correlations in multimode fibers for pulse delivery}
	
	\author{Wen Xiong}
		\affiliation{Department of Applied Physics, Yale University, New Haven, Connecticut 06520, USA}
	\author{Chia Wei Hsu} 
		\affiliation{Department of Applied Physics, Yale University, New Haven, Connecticut 06520, USA}
	\author{Hui Cao} 
		\email{hui.cao@yale.edu}
		\affiliation{Department of Applied Physics, Yale University, New Haven, Connecticut 06520, USA}
	
\begin{abstract}
Long-range speckle correlations play an essential role in wave transport through disordered media, but have rarely been studied in other complex systems. Here we discover spatio-temporal intensity correlations for an optical pulse propagating through a multimode fiber with strong random mode coupling. Positive long-range correlations arise from multiple scattering in fiber mode space and depend on the statistical distribution of arrival times. By optimizing the incident wavefront of a pulse, we maximize the power transmitted at a selected time, and such control is significantly enhanced by the long-range spatio-temporal correlations. We provide an explicit relation between the correlations and the enhancements, which closely agrees with experimental data. Our work shows that multimode fibers provide a fertile ground for studying complex wave phenomena, and the strong spatio-temporal correlations can be employed for efficient power delivery at a well-defined time.
\end{abstract}
\maketitle

Coherent transport of classical and quantum waves in disordered media exhibits long-range correlations, which exist in space, angle, frequency, time, and polarization~\cite{Akkermans07, Sheng06, 1987_Stephen_PRL, Feng88, Mello88, 2000_Sebbah_PRE, 2002_Sebbah_PRL, Garcia02, Skipetrov04, 2004_Chabanov_PRL, 2010_Wang_PRB, 2014_Hildebrand_PRL, 2015_Dogariu_PR, 2017_Riboli_PRL, 2018_Starshynov_PRX}. Such correlations, resulting from crossings of wave paths, are responsible for the formation of highly transmitting channels in diffusive systems \cite{Dorokhov84,1994_Nazarov_PRL,Geradin14,Sarma16}. In the frequency domain, long-range correlations enable broad-band enhancement of transmission through disordered media by wavefront shaping \cite{Hsu15}. Spatially, long-range correlations significantly increase the efficiency of wave focusing to a target of size much larger than the wavelength in strongly scattering media \cite{Hsu17}. 

From the aspect of scattering, a multimode fiber (MMF) with strong mode mixing shares similarities with a disordered medium. Inherent imperfections and environmental perturbations introduce random mode coupling in an MMF, and its effect grows with the length of the fiber~\cite{2018_Xiong_LSA, 2018_Chiarawongse_NJP}. Such coupling can be regarded as scattering in the fiber mode space, leading to energy transfer from the input mode to the other transverse modes. An MMF has a significant difference from the disordered medium: negligible reflection and low propagation loss leading to near-unity transmission. For a continuous wave input, energy conservation dictates that the intensity increase in one mode must be accompanied by intensity decreases in other modes, resulting in a negative correlation among the transmitted spatial modes, similar to those found in weak-scattering (ballistic) systems and chaotic cavities~\cite{Garcia02, 2010_Dietz_PLB, 2016_Gehler_PRB}. If the MMF has a large number of modes, such {\it static} correlations are very weak. When the input is a short pulse, however, energy is no longer conversed at any particular time, and correlations may be modified and become time-dependent. Nevertheless, little is known about such {\it dynamic} correlations in MMFs.

In this work, we discover spatio-temporal correlations in MMFs with strong random mode mixing. Experimentally, we find that for a short pulse input, the transmitted intensities in different spatial channels are generally positively correlated at a given arrival time. The correlations are enhanced at arrival times away from the center of the transmitted pulse, which we attribute to the reduced number of propagation paths at early or late arrival times. Such long-range dynamic correlations in an MMF are distinct from and can be much stronger than those predicted to exist in a random scattering medium~\cite{Skipetrov04}.

The spatio-temporal correlations play a crucial role in the coherent control of short pulses transmitting through an MMF. The positive correlations among spatial channels enable a global enhancement of transmitted energy at a selected arrival time by shaping the incident wavefront. Experimentally, we achieve a higher enhancement when the target time is before or after the mean arrival time, as a result of stronger long-range correlations. Theoretically, we provide a quantitative relation between spatio-temporal correlations and the time-dependent enhancement of transmitted power, which agrees well to our experimental data.
Our results show that the maximal power that can be delivered through an MMF at a well-defined time is much higher than what is achievable without long-range correlations. This discovery is important to MMF applications such as telecommunication \cite{Richardson13}, fluorescence endoscopy \cite{Cizmar12, Ploschner15, Morales-Delgado15}, nonlinear microscopy \cite{Brasselet11} and fiber amplifiers \cite{Bai11, Florentin17, Wright17}, in which ultrashort pulses are deployed for energy delivery. 
  
\begin{figure*}[t]
	\includegraphics[width=1.6\columnwidth,keepaspectratio,clip]{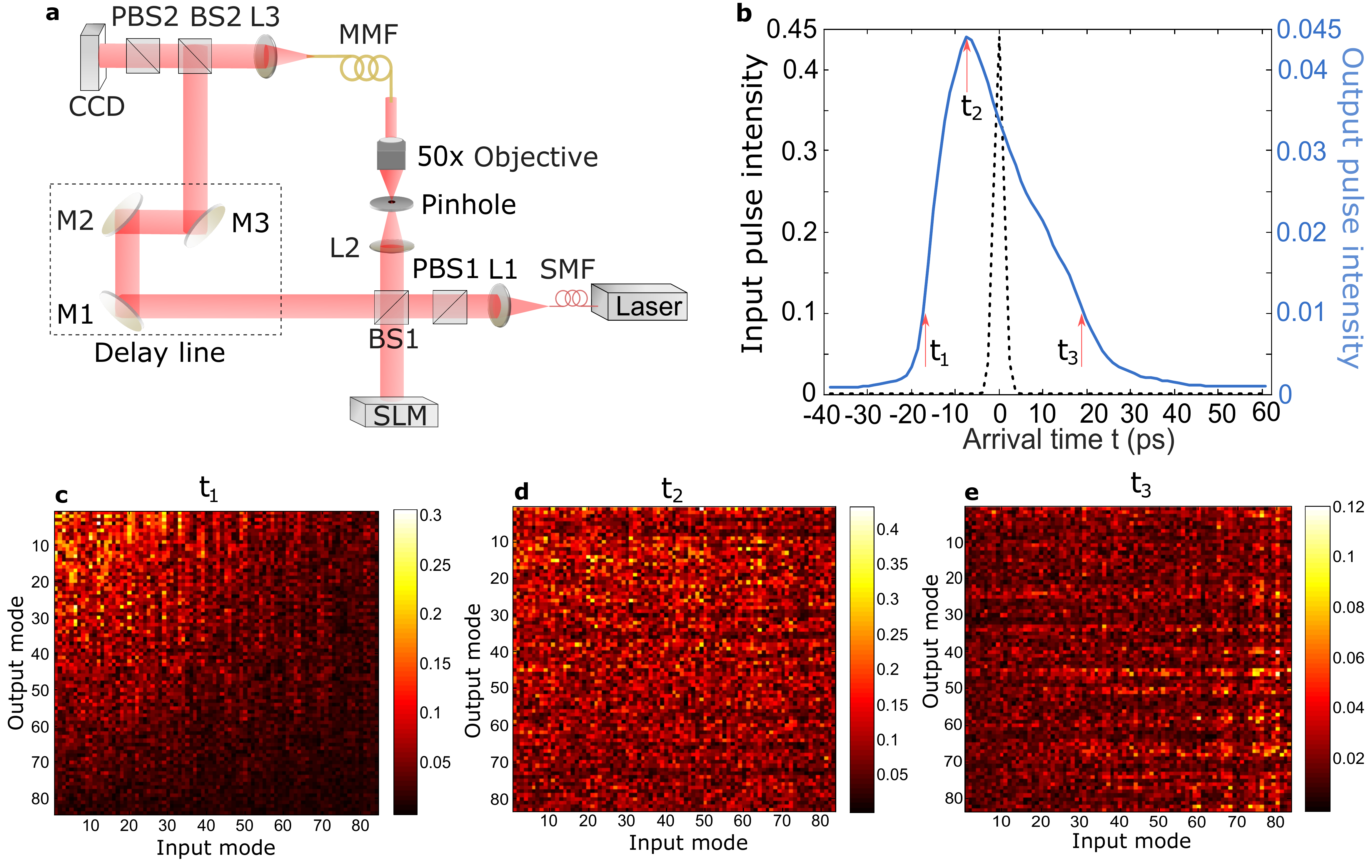}
	\caption{\textbf{Time-dependent transmission matrices of a multimode fiber (MMF).} \textbf{a}, Schematics of experimental setup for both transmission matrix measurement and wavefront shaping. A laser beam with tunable frequency is collimated, and its horizontal polarization is selected and split into two arms, with one being the reference and the other propagating through the MMF after reflecting off a spatial light modulator (SLM). The SLM is demagnified and imaged onto the MMF facet. Light transmitted through the MMF is recombined with the reference plane wave, and its horizontal polarization is imaged onto a CCD camera. The path lengths of the two arms are matched by tuning the delay line formed by mirrors M1--M3.
	L, lens; BS, beam splitter; PBS, polarizing beam splitter. 
	 \textbf{b}, Temporal shapes of the input pulse (black dashed line, left axis) and the output pulse averaged over many random spatial inputs (blue solid line, right axis). The two curves are normalized to have the same area.
	 \textbf{c-e}, Magnitudes of the measured time-dependent transmission matrices at three arrival times (marked by red arrows in \textbf{b}), showing strong mode mixing in the fiber. The transmission matrices are measured in ${\bf k}$ space at input and real space ${\bf r}$ at output, and subsequently converted to the fiber mode basis.}
	\label{fig: figure1}
\end{figure*}

We start with the known static correlations in disordered media. For a monochromatic wave with any given input wavefront, the intensities $I({\bf r})$ at different output positions ${\bf r}$ are correlated~\cite{2002_Sebbah_PRL, Cwilich06, 2008_Yamilov_PRB}
\begin{equation}
\label{eq:C_spatial}
\frac{\langle I({\bf r})I({\bf r}+\Delta{\bf r})\rangle}{\langle I({\bf r})\rangle \langle I({\bf r}+\Delta{\bf r})\rangle} - 1 = F(\Delta{\bf r})\tilde{C}_1+\tilde{C}_2, 
\end{equation}
where $\langle \cdots \rangle$ denotes ensemble average. The constant $\tilde{C}_1$ gives the strength of short-range correlation, as the normalized function $F(\Delta{\bf r})$ decays to zero at the distance $|\Delta{\bf r}|$ larger than the speckle size \cite{Goodman_book}. The constant $\tilde{C}_2$ represents the long-range correlation that results from path crossings~\cite{Akkermans07,Feng88} and is independent of distance (see more details in Supplementary, Section I). 

Even though Eq.~\eqref{eq:C_spatial} originates from wave transport in disordered media, it has been derived mathematically on fully general ground~\cite{Mello90,Cwilich06}, with the only assumption of {\it isotropy}, namely, all channels are fully mixed and are statistically equivalent.
An MMF with strong and random mode mixing can also exhibit isotropy and therefore follow Eq.~\eqref{eq:C_spatial}.
However, for continuous wave, the correlations associated with an MMF is somewhat trivial, as $\tilde{C}_2 \approx -1/N$ is small ($N$ is the number of fiber modes) and follows directly from energy conservation. 

Pulsed inputs introduce time dependences and non-trivial magnifications to the correlations in MMFs.
We consider correlations of the transmitted intensity $I({\bf r}, t)$ between different output positions ${\bf r}$ and ${\bf r} + \Delta {\bf r}$ at arrival times $t$ and $t'$,
\begin{equation}
\label{eq:C_spatiotemporal}
C(\Delta{\bf r}, t,t') \equiv \frac{\langle I({\bf r}, t)I({\bf r}+\Delta{\bf r}, t')\rangle}{\langle I({\bf r}, t)\rangle \langle I({\bf r}+\Delta{\bf r}, t')\rangle} - 1.
\end{equation}
In the $t=t'$ case, when $C(\Delta{\bf r}, t,t)$ is positive, the transmitted power at time $t$ can be efficiently enhanced by wavefront shaping, since enhancing the intensity at one position will simultaneously enhance the intensities at other positions. If the time-dependent transmission matrix at arrival time $t$ is sufficiently isotropic, we expect the same structure as Eq.~\eqref{eq:C_spatial}.
When $t \neq t'$, the correlation governs the transmitted intensities at time $t$ when the transmission at time $t'$ is modified by changing the incident wavefront; therefore it is related to the temporal shape of the output pulse when the transmitted power is optimized at a given time. 

\begin{figure*}[t]
	\includegraphics[width=1.8\columnwidth,keepaspectratio,clip]{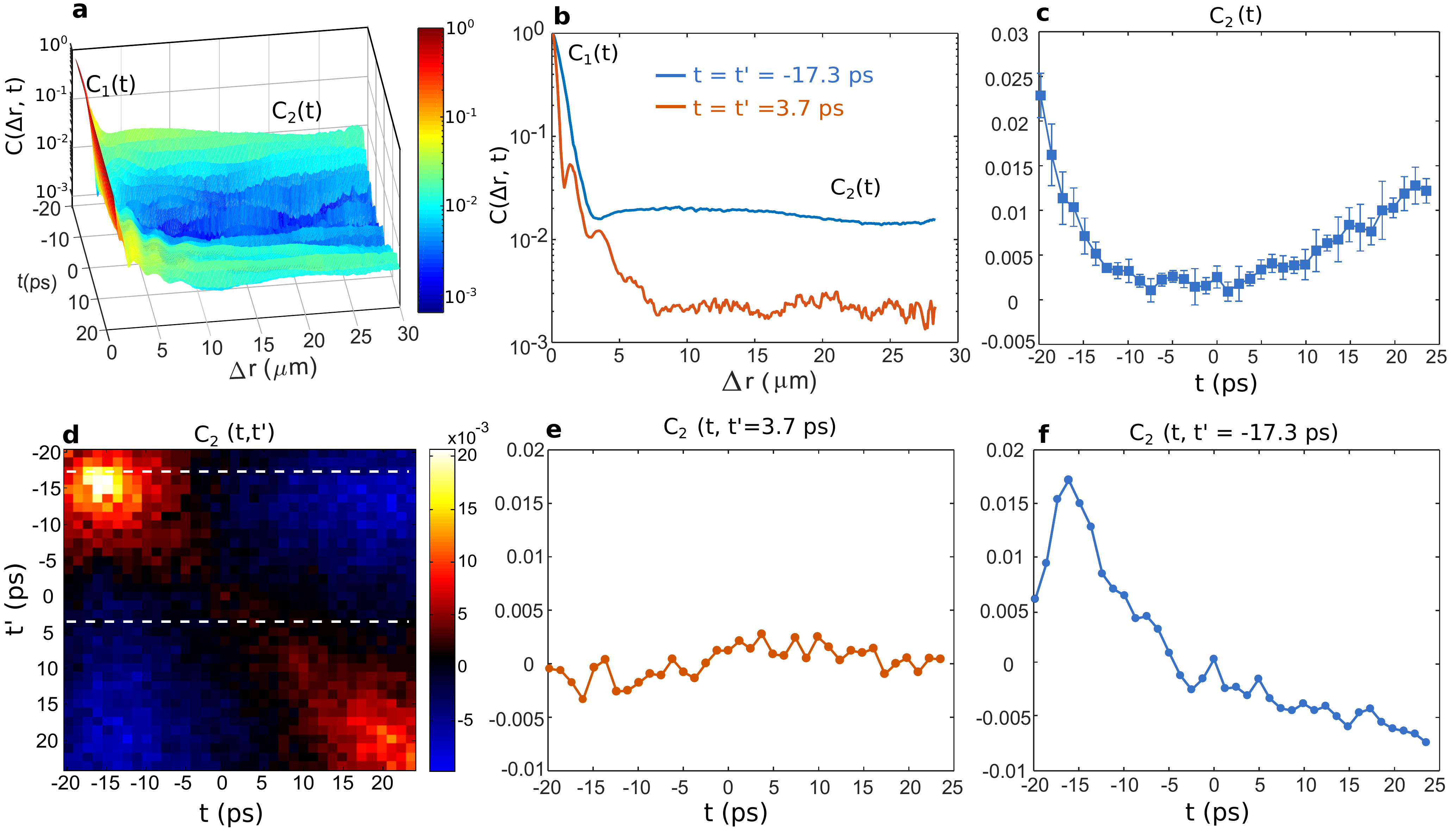}
	\caption{\textbf{Spatio-temporal correlations in MMF with strong random mode mixing.}
	\textbf{a}, Intensity correlations $C(\Delta r, t) \equiv C(|\Delta{\bf r}|, t,t'=t)$, revealing a short-range component $C_1(t) \approx 1$ at spatial distance within one speckle ($\Delta r \lesssim 3$ $\mu$m) and a long-range component $C_2(t)$ that persists at large distance.
	\textbf{b}, Two cross sections of $C(\Delta r, t)$ at arrival times $t = -17.3$ ps and $t = 3.7$ ps.
	\textbf{c}, Time dependence of the long-range component $C_2(t)$, averaging over $\Delta r$ for $\Delta r > 5$ $\mu$m.
	\textbf{d}, Long-range correlations $C_2(t,t')$ between spatio-temporal speckles at different arrival times $t$ and $t'$.
	\textbf{e-f}, Cross sections of $C_2(t,t')$ at $t' = 3.7$ ps and $-17.3$ ps (marked by white dashed lines in \textbf{d}).} 
	\label{fig: figure2}
\end{figure*}

To characterize such spatio-temporal correlations, we measure the transmission matrix of an MMF with strong mode mixing. We use an off-axis holographic setup schematically shown in Fig.~\ref{fig: figure1}\textbf{a}. A spatial light modulator (SLM; Hammamatsu X10468) scans the incident angle of a laser beam (Aglient  81940A) onto the MMF, to excite different spatial modes with horizontal polarization. The plane wave of the reference arm and the light transmitted through the fiber interfere to form fringes on the camera, from which we extract the horizontally polarized transmitted field. We use an one-meter-long 0.22-NA graded-index fiber with a core radius of 50 $\mu$m and 84 guided modes per polarization. To introduce strong mode mixing to such a short fiber, we use clamps to create micro-bendings. The path lengths of the two arms are matched so the mean arrival time of the pulse (relative to the reference) is zero. The spectral correlation width of the fiber is 0.20 nm at the wavelength of 1550 nm. We measure the field transmission matrices over a wavelength range of 6.4 nm with the step of 0.04 nm.
We then perform a Fourier transform to obtain the time-dependent transmission matrices $u(t)$ relating the incident wavefront $|\psi_{\rm in}\rangle$ to the transmitted wavefront $|\psi_{\rm out} (t)\rangle = u(t) |\psi_{\rm in}\rangle$ at different arrival times $t$, considering a Gaussian transform-limited input pulse centered at wavelength 1550 nm with a full width at half maximum (FWHM) of 2.0 nm (temporal FWHM = 2.6 ps).

The mean temporal shape of the transmitted pulse, obtained by averaging over many random spatial profiles of incident light, is shown in Fig.~\ref{fig: figure1}\textbf{b} together with the input pulse.  The output pulse is significantly broadened due to mixing between modes that propagate at different group delays. The strong random mode coupling is evident from the magnitude of the time-dependent transmission matrix, shown in Fig.~\ref{fig: figure1}\textbf{d} for central arrival time; no matter which mode is launched at the input, light is scattered to all spatial modes at the output. Higher-order modes have slightly lower magnitudes due to mode-dependent loss in the fiber. The transmission matrix at early (late) arrival time in Fig.~\ref{fig: figure1}\textbf{c} (Fig.~\ref{fig: figure1}\textbf{e}) has higher concentrations of lower-order (higher-order) modes, which have shorter (longer) transit times through the fiber.

We calculate the spatio-temporal correlations $C(\Delta{\bf r}, t,t')$ from the measured time-dependent transmission matrices, replacing the ensemble average in Eq.~\eqref{eq:C_spatiotemporal} with an average over random input spatial profiles. Figure~\ref{fig: figure2}\textbf{a}-\textbf{b} plot $C(\Delta r, t) \equiv C(|\Delta{\bf r}|, t,t'=t)$ and two cross sections of it along $\Delta r$ at $t = -17.3$ ps and $t = $ 3.7 ps. We observe a short-range correlation that starts from one and vanishes at the speckle size of about 3 $\mu$m, beyond which we see a long-range correlation that is approximately constant with respect to $\Delta r$. This indicates $C(\Delta r, t) = F(\Delta r) C_1(t) + C_2(t)$, consistent with Eq.~\eqref{eq:C_spatial}. Figure~\ref{fig: figure2}\textbf{c} shows the arrival-time dependence of the long-range correlation $C_2(t)$; it is small at the central arrival time but increases toward early or late arrival times. 

\begin{figure*}[t]
	\includegraphics[width=1.6\columnwidth,keepaspectratio,clip]{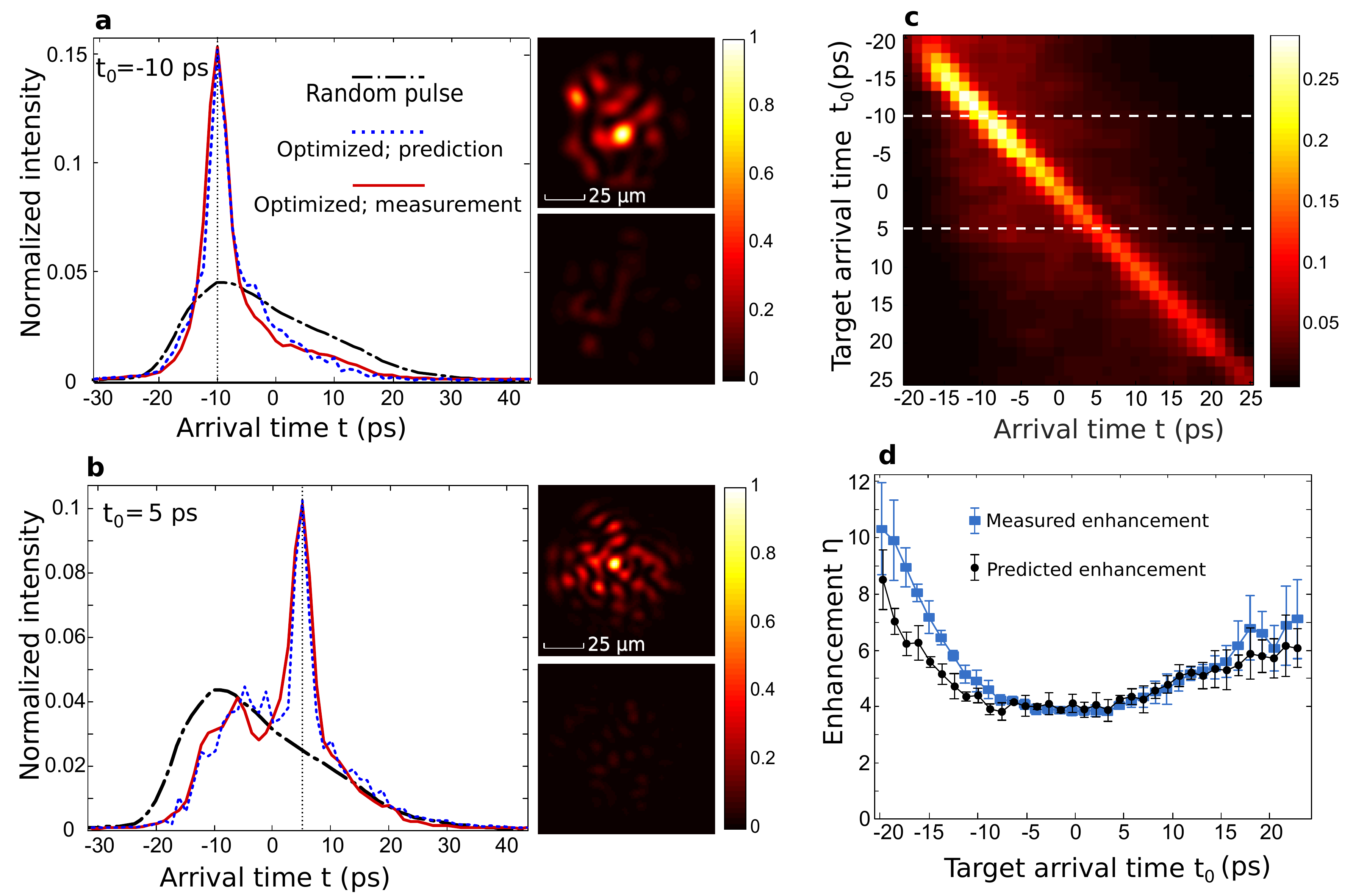}
	\caption{\textbf{Enhancing transmitted power at selected time.}
	\textbf{a-b,} Temporal shapes of the output pulse when the spatially integrated intensity (power) is optimized at arrival time (\textbf{a}) $t_0$ = -10 ps and (\textbf{b}) $t_0$ = 5 ps. Red solid lines are the measured pulse shapes with optimized input wavefronts, the blue dotted lines are predicted from measured correlations $C_2(t,t')$. They exhibit strong enhancement compared to the output pulse from a random wavefront (black dash-dotted lines).
	The insets are spatial intensity patterns at $t_0$ for the optimized wavefront (upper panel) and a random wavefront (lower panel).
	\textbf{c}, Temporal shapes of pulses optimized at different $t_0$, ranging from -20 ps to 25 ps. The target time of (\textbf{a}) and (\textbf{b}) are marked by the white dashed lines.
	\textbf{d}, Enhancement factor $\eta$ of the transmitted power at the target arrival time $t_0$. Blue squares: measured enhancement. Black circles: enhancement predicted from $C_2(t)$. Error bars indicate standard deviation among different measurements.
	}
	\label{fig: figure3}
\end{figure*}

The time dependence of $C_2(t)$ can be understood through the distribution of arrival times, which is given by the number of optical paths in the fiber for a given arrival time. Conceptually, we expect that if there is only one path that can arrive at time $t$, then the output intensities $I({\bf r}, t)$ at different positions ${\bf r}$ must be fully correlated: varying the incident wavefront can only change how much light is coupled into that one path, which will increase or decrease $I({\bf r}, t)$ at all positions in the same way. Therefore we expect the fewer paths at time $t$, the stronger $C_2(t)$ would be. This is analogous to the disordered medium where the static $C_2$ is inversely proportional to the number of high-transmission channels~\cite{Akkermans07}.
The distribution of arrival times is captured by the average shape of the output pulse, shown in Fig.~\ref{fig: figure1}\textbf{b}. Indeed, the large $C_2(t)$ at early and late arrival times is consistent with the lower number of paths at such times.

Long-range correlations between far-away speckles exist not only between speckles at the same arrival time, but also between speckles at different arrival times. This is quantified by $C(\Delta{\bf r}, t,t')$ as defined in Eq.~\eqref{eq:C_spatiotemporal}. At large $|\Delta{\bf r}|$, this quantity again becomes independent of $|\Delta{\bf r}|$ and approaches the asymptotic value $C_2(t,t')$. In Fig.~\ref{fig: figure2}\textbf{d}, we show $C_2(t, t')$ for $t$ and $t'$ from $-20$ ps to $25$ ps. The long-range correlations are positive close to the diagonal, namely close to the $C_2(t)$ discussed earlier. When $t$ and $t'$ are far apart, however, the long-range correlations become negative. Fig.~\ref{fig: figure2}\textbf{e-f} show two cross-sections. Near the central arrival time ($t' = 3.7$ ps), $C_2(t, t')$ is close to zero at all $t$ (Fig.~\ref{fig: figure2}\textbf{e}). Meanwhile, at $t' = -17.3$ ps (Fig.~\ref{fig: figure2}\textbf{f}), $C_2(t, t')$ peaks at $t \approx t'$ and decays away from it, eventually becoming negative. Such a negative correlation is a result of the conservation of transmitted pulse energy, which requires an increase of spatially integrated intensity (power) at arrival time $t=t'$ to be compensated by a decrease of power at other arrival times.

By shaping the incident wavefront with the SLM~\cite{2012_Mosk_nphoton, 2017_Rotter_RMP}, we can enhance the total (spatially integrated) intensity at a target arrival time and compensate for the strong modal dispersion in the fiber. The positive spatio-temporal correlations $C_2(t)$ at early or late arrival times will lead to a higher achievable enhancement at such times. Given a spatially shaped incident wavefront $|\psi_{\rm in}\rangle$, the total output intensity at arrival time $t_0$ is $\langle \psi_{\rm out} (t_0) | \psi_{\rm out} (t_0) \rangle$ = $\langle \psi_{\rm in} | u^\dagger(t_0) u(t_0) | \psi_{\rm in} \rangle $. As this is the expectation value of a Hermitian matrix $u^\dagger(t_0) u(t_0)$, the global optimum, which determines the maximum power that can be delivered at time $t_0$, is given by the largest eigenvalue of $u^\dagger(t_0) u(t_0)$, and the corresponding eigenvector is the desired incident wavefront~\cite{Hsu15}.
Experimentally, we determine such optimal transmission channels from the measured time-dependent transmission matrices, and then generate the desired wavefront with the same setup, using computer-generated phase holograms to simultaneously modulate the phase and amplitude profiles~\cite{2007_Arrizon_JOSAA}. By scanning the wavelength and Fourier transforming the spectral  measurements to time domain, we obtain the spatially integrated temporal pulse shapes of such optimal transmission channels.

The pulses optimized for arrival times $t_0 = -10 $ ps and $t_0 = 5$ ps are shown in Fig.~\ref{fig: figure3}\textbf{a-b} (red solid curve), in comparison to the averaged pulse of random spatial inputs (black dash-dotted curve). The sharp peak at the selected arrival time, as marked by the vertical black dotted line, illustrates that the transmitted power can be effectively enhanced at different target times, even in the presence of strong modal dispersions in the fiber. The spatial intensity patterns of the optimized pulse and the random pulse at the target arrival times are also shown in Fig.~\ref{fig: figure3}\textbf{a-b}. All of the spatial channels are enhanced, which is distinct from spatiotemporal focusing~\cite{Katz11,McCabe11,Aulbach11,Mounaix16, Papadopoulos12} where only one speckle is enhanced.  Figure~\ref{fig: figure3}\textbf{c} plots the pulse shapes optimized for different arrival times from $t_0$ = -20 ps to 25 ps. The peak follows the target time $t_0$, and notably, the background also shifts with the target time. The background change is a result of long-range spatio-tempral correlations $C_2(t, t')$. Namely, the negative correlations for well-separated times $t$ and $t'$ ($t'= t_0$) shift the background toward the target time $t_0$. 

To evaluate the effectiveness of the transmitted power optimization, we define an enhancement factor $\eta(t_0) \equiv I_{\rm enh}(t_0)/I_{\rm random}(t_0)$, where $I_{\rm enh}(t_0)$ and $I_{\rm random}(t_0)$ are the spatially integrated intensities of the optimized pulse and the random pulse at the target time $t_0$. We plot the measured enhancement factor $\eta$ (blue square) in Fig.~\ref{fig: figure3}\textbf{d}. The standard deviation of the enhancement between measurements on four different days is shown by the error bars. The deviation is larger at early or late arrival times as the weaker pulse intensities there lead to smaller signal-to-noise ratio. The average enhancement is about 4 times around the central arrival time, which is what one expects through the quarter-circle law for the singular values of a square random matrix with uncorrelated elements~\cite{Marcenko67}. At early or late arrival times, we achieve power enhancements much larger than 4; such increase is consistent with the large long-range correlations $C_2(t)$ that we observed (Fig.~\ref{fig: figure2}\textbf{c}). 

Finally, we provide a quantitative connection between the long-range spatio-temporal correlations $C_2(t)$ and the enhancement factor $\eta(t_0)$ of transmitted power at arrival time $t_0$. We use a heuristic model similar to that employed in Ref.~\cite{Hsu17}, capturing the correlations between output channels at arrival time $t_0$ through a reduction in the effective number of output channels.
Specifically, we consider an effective random matrix with $N$ input channels and $N^{\rm (eff)}(t_0)$ output channels, and we consider all  elements of this matrix to be identically independently distributed.
The enhancement $\eta$ is determined by the largest eigenvalue, which is related to the spread of the eigenvalues characterized by the eigenvalue variance.
As detailed in the Supplementary Section I, the normalized eigenvalue variance associated with the reduced matrix is given by the Mar\v chenko--Pastur distribution~\cite{Marcenko67} to be $N/N^{\rm (eff)}(t_0)$, while that associated with the actual time-resolved transmission matrix is $1 + N C_2(t_0)$.
Therefore, we choose
\begin{equation}
\label{eq:M2_eff}
N^{\rm (eff)}(t_0) = \frac{N}{1 + N C_2(t_0)}
\end{equation}
to match the two corresponding eigenvalue variances.
This relation quantifies how long-range correlations effectively reduce the number of output channels.
The enhancement is the normalized maximal eigenvalue, which for an uncorrelated matrix is $\tau_{\rm max}/\bar{\tau} = \left(1 + \sqrt{N/N^{\rm (eff)}(t_0)} \right)^2$ (Ref.~\cite{Marcenko67}).
Inserting Eq.~\eqref{eq:M2_eff}, we obtain a simple equation
\begin{equation}
%\eta(t) = 2 + N C_2(t) + 2\sqrt{1 + N C_2(t)}
\eta(t_0) = \left(1 + \sqrt{1 + N C_2(t_0)} \right)^2
\label{eq: enhance-correlation}
\end{equation}
that relates the maximal enhancement to the long-range spatio-temporal correlations. 

In Fig.~\ref{fig: figure3}\textbf{d}, we compare the measured enhancement to the enhancement predicted through the measured $C_2(t)$ via Eq.~\eqref{eq: enhance-correlation} (black circles).  Overall, the two curves agree well, especially around the central arrival time. Some differences at early or late arrival times may be due to the fact that the time-dependent transmission matrix is not as isotropic as that at the central time (as shown in Fig.~\ref{fig: figure1}\textbf{c-e}). We further generalize the relationship in Eq.~(\ref{eq: enhance-correlation}) to predict the whole output pulse (both the peak at the target time and the background) via $C_2(t, t')$ (see Supplementary, Section I). In Figs.~\ref{fig: figure3}\textbf{a-b}, we plot the predicted temporal shapes (blue dotted curves) of the optimized pulses with $t_0 = -10$ ps and $t_0 = 5$ ps on top of the measured pulse shapes. The theoretical model well captures the temporal transmission of the MMF.  

Speckle correlations have been studied extensively in scattering media, but there are few observations in other complex photonic systems. We observe spatio-temporal correlations in a multimode fiber with strong mode mixing when a pulse propagates through the fiber. The correlations not only determine the effectiveness of enhancing the transmitted power at a target time, but also capture the temporal shape of the resulting pulse. Enhancing the transmitted power in time can be utilized in many fiber applications from communication to imaging. In comparison, while principal modes~\cite{Carpenter15,Xiong16,Xiong17} and super-principal modes~\cite{Ambichl17} have been demonstrated recently to maintain the shape of an incident pulse, they have limited spectral bandwidth and work only for long pulses. For the broad-band pulses studied here, selective excitation of time-dependent transmission eigenchannels with extreme eigenvalues is the most efficient way to maximize the transmitted power at the target time (see Supplementary, Section II). 

The authors acknowledge Stefan Rotter, Hasan Yilmaz, Tsampikos Kottos, Douglas Stone, and Alexandre Aubry for stimulating discussions. This work is supported by US National Science Foundation under the Grant No. ECS-1809099.

\clearpage


\section*{Supplementary material}

# Spatio-temporal correlations in multimode fibers for pulse delivery

Wen Xiong, Chia Wei Hsu, and Hui Cao*
*Department of Applied Physics, Yale University, New Haven, Connecticut 06520, USA*## I. CORRELATIONS AND TEMPORAL ENHANCEMENT

In this section, we first review the mathematics behind the known static correlations in disordered media, and then discuss their consequences for the spatio-temporal correlations in multimode fibers. We then introduce our model for temporal control.

### A. Correlations and transmission eigenvalues

The propagation of monochromatic light through a system with $N$ input and $N$ output channels is described by an $N$-by-$N$ transmission matrix $u$, where the matrix element $u_{ba}$ is the flux-normalized field transmission coefficient from input channel $a$ to transmitted channel $b$. We denote $U_{ba} \equiv |u_{ba}|^2$.

We consider the singular-value decomposition $u = W\sqrt{\tau}V^\dagger$; here $W$ and $V$ are $N$-by-$N$ unitary matrices, and $\tau$ is a diagonal matrix whose elements $\{\tau_n\}_{n=1}^N$ are the eigenvalues of $u^\dagger u$.

For disordered media, the intensity correlation between the transmitted speckles is defined in channel space as

$$C_{aa',bb'} \equiv \frac{\langle U_{ba} U_{b'a'} \rangle}{\langle U_{ba} \rangle \langle U_{b'a'} \rangle} - 1 \quad \text{(S1)}$$

where $\langle \cdots \rangle$ denotes ensemble average over different disorder realizations. If the matrices $u$ are *isotropic*, the correlations must take on the form [1, 2]

$$C_{aa',bb'} = \delta_{aa'}\delta_{bb'}C_1 + (\delta_{aa'} + \delta_{bb'})C_2 + C_3, \quad \text{(S2)}$$

where the constants $C_1$, $C_2$, $C_3$ are commonly referred to as the magnitudes of the short-range, long-range, and infinite-range correlations. Typically $C_1 \gg C_2 \gg C_3$.

Mathematically, the transmission matrices $u$ are "isotropic" when the matrices $W$ and $V$ are sampled uniformly and independently from the space of all random unitary matrices in the ensemble average. Physically, isotropy means that all input modes and all output modes are fully mixed, and that all modes are statistically equivalent.

Spatially, a similar structure emerges for intensity correlation function. Denote $I(\mathbf{r}_b, \mathbf{r}_a)$ as the transmitted intensity at position $\mathbf{r}_b$ on the back surface given a point-source excitation at position $\mathbf{r}_a$ on the front surface.

---

*Electronic address: hui.cao@yale.edu

For isotropic disordered media, the correlation between $I(\mathbf{r}_b, \mathbf{r}_a)$ and $I(\mathbf{r}_{b'}, \mathbf{r}_{a'})$ takes on the form [3, 4]

$$C(\mathbf{r}_{a'} - \mathbf{r}_a, \mathbf{r}_{b'} - \mathbf{r}_b) = [F(\mathbf{r}_{a'} - \mathbf{r}_a)F(\mathbf{r}_{b'} - \mathbf{r}_b)]C_1 \\ + [F(\mathbf{r}_{a'} - \mathbf{r}_a) + F(\mathbf{r}_{b'} - \mathbf{r}_b)]C_2 + C_3, \quad \text{(S3)}$$

where the coefficients $C_1$, $C_2$ and $C_3$ are the same as in Eq. (S2). $F(\Delta \mathbf{r})$ is a function that decays from one when $\Delta \mathbf{r} = 0$ to zero when $|\Delta \mathbf{r}|$ is large. Given a fixed input such that $\mathbf{r}_{a'} = \mathbf{r}_a$, we obtain Eq. (1) in the main text, where we define $\tilde{C}_1 \equiv C_1 + C_2 \approx C_1$ and $\tilde{C}_2 \equiv C_2 + C_3 \approx C_2$.

The statistics of the transmission eigenvalues uniquely determines the magnitudes of the correlations. It was shown rigorously that [1, 2]

$$\begin{aligned} C_1 &= \frac{N^2(N^2+1)}{(N^2-1)^2}\left(\frac{\langle\alpha^2\rangle}{\langle\alpha\rangle^2} - \frac{2N}{N^2+1}\frac{\langle\alpha_2\rangle}{\langle\alpha\rangle^2}\right), \\ C_2 &= \frac{N^2(N^2+1)}{(N^2-1)^2}\left(\frac{\langle\alpha_2\rangle}{\langle\alpha\rangle^2} - \frac{2N}{N^2+1}\frac{\langle\alpha^2\rangle}{\langle\alpha\rangle^2}\right), \\ C_3 &= C_1 - 1, \end{aligned} \quad \text{(S4)}$$

where $\alpha \equiv \sum_{n=1}^N \tau_n$ and $\alpha_2 \equiv \sum_{n=1}^N \tau_n^2$.

Even though the preceding results were initially obtained in the context of monochromatic light propagation through disordered media, the mathematical framework is very general. In fact, Eqs. (S2)–(S4) are valid as long as the matrices $u$ are isotropic [2, 3]. In particular, the propagation of a *pulse* through a *multimode fiber* will also satisfy Eqs. (S2)–(S4) if the time-resolved transmission matrix $u(t)$ at arrival time $t$ is sufficiently isotropic.

### B. Effective model for temporal enhancement

As described in the main text, the largest-possible enhancement of the spatially integrated intensity at a target arrival time $t_0$ is given by the largest eigenvalue of $u^\dagger(t_0)u(t_0)$. Eq. (S4) relates the spread of eigenvalues to the magnitudes of the correlations. Here, we provide a heuristic model to relate the maximal eigenvalue to the spread of eigenvalues and then to the correlations.

We model the eigenvalues $\{\tau_n(t_0)\}_{n=1}^N$ associated with the $N$-by-$N$ matrix $u(t_0)$ using the eigenvalues $\{\tau_n^{(\mathrm{eff})}(t_0)\}_{n=1}^N$ associated with an effective $N^{(\mathrm{eff})}(t_0)$-by-$N$ matrix $u^{(\mathrm{eff})}(t_0)$ with uncorrelated matrix elements. The idea, as first introduced in Ref. [5], is to capture the effects of correlations in $u(t_0)$ by resizing the number of output channels from $N$ to $N^{(\mathrm{eff})}(t_0)$. As the magnitude of correlations depends on the arrival time (see Fig. 2c in

the main text), the effective number $N^{(\text{eff})}(t_0)$ will also depend on $t_0$.

Given a large $M$-by-$N$ random matrix $u$ with uncorrelated elements, the eigenvalues of $u^\dagger u$ are given by the Marčhenko–Pastur distribution [6], with the largest eigenvalue being $\tau_{\max} = (1+\sqrt{N/M})^2 \bar{\tau}$ and the eigenvalue variance being $\text{var}(\tau) \equiv \overline{\tau^2} - \bar{\tau}^2 = (N/M)\bar{\tau}^2$. Therefore, the normalized maximal eigenvalue is directly related to the spread of eigenvalues, as

$$\frac{\tau_{\max}}{\bar{\tau}} = \left(1 + \sqrt{\frac{\text{var}(\tau)}{\bar{\tau}^2}}\right)^2. \quad \text{(S5)}$$

Here, we use overhead bars to denote averaging over the $N$ eigenvalues.

The spread of eigenvalues is related to the magnitudes of the correlations through Eq. (S4). Specifically, we obtain

$$\frac{\langle \text{var}(\tau)\rangle}{\langle \bar{\tau}\rangle^2} = \left(1 - \frac{1}{N^2}\right)(1 + NC_2 + C_3) \approx 1 + NC_2 \quad \text{(S6)}$$

by writing $\text{var}(\tau) = (\alpha_2/N) - (\alpha^2/N^2)$ and solving for $\langle \alpha_2\rangle/\langle \alpha\rangle^2$ and $\langle \alpha^2\rangle/\langle \alpha\rangle^2$ in Eq. (S4).

Inserting Eq. (S6) into Eq. (S5), we get

$$\eta(t_0) = 2 + NC_2(t_0) + 2\sqrt{1 + NC_2(t_0)}, \quad \text{(S7)}$$

which is Eq. (4) in the main text. This expression concisely relates the long-range spatio-temporal correlations $C_2(t)$ to the maximal power enhancement at a targeted arrival time $t_0$. It has two underlying assumptions, the first being the isotropy of $u(t_0)$ [which underlies Eq. (S4) and Eq. (S6)], the second being the heuristic modeling of $\{\tau_n\}$ using a resized uncorrelated matrix [which underlies Eq. (S5)].

### C. Model for optimized pulse shape

Experimentally, we also measure the long-range correlations between speckles at different arrival times. Such information can be used to predict the pulse shape (output power as a function of $t$) when the input wavefront has been optimized to enhance the power at arrival time $t_0$. We construct a heuristic model based on two observations. First, the normalized power $\eta(t, t_0)$ should be identical to Eq. (S7) when $t = t_0$. Second, when $t$ is far from $t_0$, the correlation $C_2(t, t_0) \approx 0$, so the power at $t$ should equal to that of a random input, giving $\eta(t, t_0) \approx 1$. To satisfy these two constraints, we propose

$$\begin{aligned}\eta(t, t_0) &= 2C_1(t, t_0) + NC_2(t, t_0) \\ &\quad + 2\sqrt{1 + NC_2(t, t_0)} - \beta[1 - C_1(t, t_0)].\end{aligned} \quad \text{(S8)}$$

$C_1(t, t_0) = 1$ for $t = t_0$, and $C_1(t, t_0) = 0$ for $t \neq t_0$. The parameter $\beta$ is chosen such that the temporally-integrated output power (pulse energy) equals that from

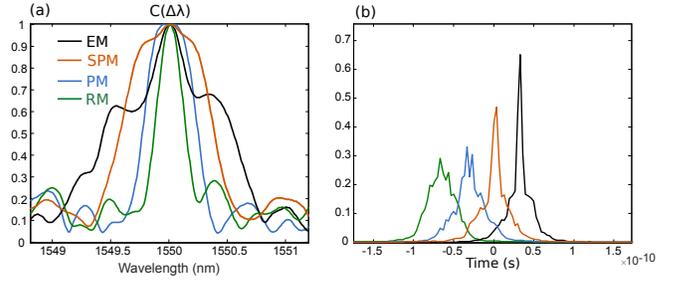

FIG. S1: (a) Spectral correlations function $C(\Delta\lambda)$ of output field pattern when the input state is the eigenvector of $u^\dagger(t_0)u(t_0)$ (EM, black), the principal mode (PM, blue), the super-principal mode (SPM, brown), and a random superposition of fiber modes (RM, green). (b) Corresponding temporal pulses at the output for the same input pulse of frequency bandwidth 10 times of the spectral correlation width of the fiber. The pulse for each input is off-set by 30 ps for clear visualization.

a random input. From numerical simulations of MMFs without loss, we find that $\beta \approx 1$. With loss, both numerical and experimental results confirm that Eq. (S8) is still an excellent model with $\beta = 1$.

### II. TEMPORAL ENHANCEMENT AND PRINCIPAL MODES

To enhance the transmitted power of a pulse at any selected time, we must overcome modal dispersions in the multimode fiber. This is demonstrated in this work by manipulating incident wavefront to control spatio-temporal interference in the fiber. Other methods based on principal modes (PMs) and super-principal modes (SPMs) also suppress modal dispersions to maintain the temporal shape of optical pulses. In this subsection, we compare these methods for broadband pulses.

The PMs are the eigenvectors of time-delay operator [7]. They represent unique input states that produce frequency-independent output field patterns to the first order of frequency variation. In the time domain, PMs can compensate for pulse stretching and distortion induced by modal dispersion of a multimode fiber. However, in the presence of strong mode coupling, the spectral bandwidth of PMs is about twice of the spectral correlation width of the mulitmode fiber, thus they are applicable only to relatively long pulses. The super-PMs are obtained by optimizing the input wavefronts to minimize spectral decorrelation of output field patterns [8]. Although they have broader bandwidth than the PMs, the improvement is merely a factor of 2 in the strong mode coupling regime. Therefore, neither PMs nor SPMs are effective in delivering short pulses considered in this work, whose frequency bandwidth is approximately 10 times of the fiber spectral correlation width, through the multimode fiber with strong mode coupling.

We compare the eigenvectors of time-delay matrix

2

(EM) to PMs and SPMs. Numerically we simulate a multimode waveguide with the concatenated model [9]. The input pulse has a Gaussian spectrum of width 10 times of the spectral correlation width of the fiber which has strong random mode coupling. We first compare the spectral correlation function of the output field patterns, $C(\Delta\lambda = \lambda - \lambda_0) = \langle\psi_{out}(\lambda_0)|\psi_{out}(\lambda)\rangle$, which characterizes how effectively the modal dispersion is suppressed [7]. Figure S1(a) plot $C(\Delta\lambda)$ for the eigenvector of $u(t_0)^\dagger u(t_0)$ (EM) with $t_0$ being the mean arrival time, the PM with the largest spectral bandwidth, the SPM with the minimum spectral decorrelation, and a random superposition of fiber modes (RM). Both PM and SPM display a plateau in the vicinity of the center wavelength (1550 nm). However, they decay faster than the EM as the wavelength moves far away from the center wavelength. Overall the EM has the least spectral decorrelation as compared to RM, PM and SPM. This is confirmed by its maximum area for spectral integration of $C(\Delta\lambda)$ in the frequency range of input pulse. Figure S1(b) compares the temporal shape of transmitted pulses for EM, PM, SPM and RM. The EM has the highest peak power, confirming it outperforms both PM and SPM for broadband pulses.


\end{document}